\documentclass[preprint,secnumarabic,showpacs,showkeys,amssymb, amsmath,nobibnotes, aps, pre]{revtex4}

\usepackage{graphicx}
\begin{document}

%\preprint{APS/123-QED}

\title{Maximum complexity distribution of a monodimensional ideal gas
out of equilibrium}
% Force line breaks with \\

\author{Xavier Calbet}
\email{xcalbet@yahoo.es}
\affiliation{
Instituto de Astrof{\'\i}sica de Canarias,\\
V{\'\i}a L\'actea, s/n,\\
E-38200 La Laguna, Tenerife, Spain.\\
}

\author{Ricardo L\'opez-Ruiz}
\email{rilopez@unizar.es}
\affiliation{
DIIS and BIFI,\\
Universidad de Zaragoza,\\
E-50009 Zaragoza, Spain.\\
}

\date{\today}

\begin{abstract}
The maximum complexity momentum distribution for an 
isolated monodimensional ideal gas out of equilibrium
is derived analytically.
In a first approximation, it consists of
a double non-overlapping Gaussian distribution. 
In good agreement with this result,
the numerical simulations of a particular isolated 
monodimensional gas, which is abruptly pushed far 
from equilibrium, 
shows the maximum complexity distribution 
in the decay of the system
toward equilibrium.  
\end{abstract}

%Use showkeys class option if keyword
\maketitle

\section{Introduction}

Boltzmann--Gibbs (BG) statistics works
perfectly for classical systems in equilibrium 
under the action of short-range forces.
But most systems in nature are out of equilibrium 
and there is no a priori reason why a particular 
phenomenon should behave according to a specific kind of statistics.
If the system finally decays toward equilibrium, then
the asymptotic long-time limit should be that of BG statistics.

Different kinds of statistics have been proposed 
to model different nonequilibrium situations.
For instance, {\it non-extensive thermostatistics} 
is based on maximizing the Tsallis 
entropy \cite{tsallis88} under different assumptions
for calculating the expectation value of the energy. 
Power law distributions are obtained by fixing the
total energy of the system in all the cases 
analyzed in Ref. \cite{ferri05}. Thus, in this scheme, 
the exponential distribution of BG statistics turns out to be a singularity that 
is recovered in the limit $q\rightarrow 1$, 
where $q$ is called the index of non-extensitivity. 
Although this type of statistics might seem a mathematical 
artifact without applications, 
several types of generalized stochastic dynamics have been recently 
constructed for which Tsallis statistics can be proved 
rigorously \cite{wilk00}. 
Also, it has been found useful in explaining many other physical 
phenomena \cite{boghosian96,reynolds03}. 
Another formalism to study out of equilibrium
situations is {\it superstatistics}.
Originally proposed by Beck and Cohen \cite{beck03}, it
deals with nonequilibrium systems with a
long-term stationary state that possess a spatio--temporally fluctuating 
intensive quantity. After averaging over the fluctuations one can obtain 
an infinite set of general statistics
called superstatistics, which constitute a superposition
of BG distributions.
Tsallis statistics is a special case of such superstatistics, and, 
in particular,
BG statistics is also recovered when $q\rightarrow 1$, where $q$ is now
a dynamical parameter with a certain 
physical interpretation.
In general, complex nonequilibrium problems may require different types of 
superstatistics \cite{beck03}.

Although all these statistical techniques for modeling  
out of equilibrium situations are becoming a well established theory,
as far as we know, there are no general laws telling us in what 
manner a system 
should relax towards equilibrium. The second law of thermodynamics 
claims that the average entropy or disorder must increase
when an isolated system tends to equilibrium but no more insight is
obtained from this postulate. In fact, this law in no way forbids 
local complexity from arising \cite{gell-mann95}.
An inspiring example is life, which can continue to exist and
replicate in an isolated system as long as internal resources last.
It could then be postulated that in an isolated system,
besides an increase in entropy, the system will try
to stay close to the maximum complexity state.
This behavior was found in Ref. \cite{calbet01} for a particular system, 
the ``tetrahedral'' gas, when complexity is defined as 
in Ref. \cite{lopez-ruiz95} 
(referred to in the literature as the LMC complexity).
Furthermore, it was established that this isolated system relaxes towards 
equilibrium by approaching {\it the maximum complexity path}.
This path is an attractive trajectory in the distribution space
connecting all the maximum complexity distributions (see Ref. \cite{calbet01}
for details).

In this paper, we perform the study of 
an isolated monodimensional ideal gas that is
initially in equilibrium, receives a strong perturbation, 
and finally freely decays towards equilibrium.
The one-particle momentum distribution is 
computationally calculated for each time during the relaxation process.
It is found that this distribution 
coincides with the maximum LMC-complexity one-particle momentum distribution.
Hence, the maximum complexity path 
in the space of one particle momentum distributions also seems to explain
the statistical evolution of this system when it approaches equilibrium.

\section{Numerical simulations of a monodimensional gas}

In order to have a graphical picture of the
processes involved, we start by describing
the numerical simulations. The gas is initially,
at time $t=0$, in equilibrium. Its
one particle momentum distribution is
described by a Gaussian or Maxwell--Boltzmann function.
At this point, two new extremely energetic particles are introduced
into the gas, forcing the gas into a far
from equilibrium state.
The system is kept isolated from then on.
It eventually relaxes again toward equilibrium showing
asymptotically another Gaussian distribution.
Most of the time, during this out of equilibrium process,
the momentum distribution function is described to a first
approximation by two Gaussian distributions.
This double Gaussian distribution coincides
with the analytically derived maximum complexity
distribution, which will be derived later.

\begin{figure}
\includegraphics[angle=-90,width=0.9\columnwidth]{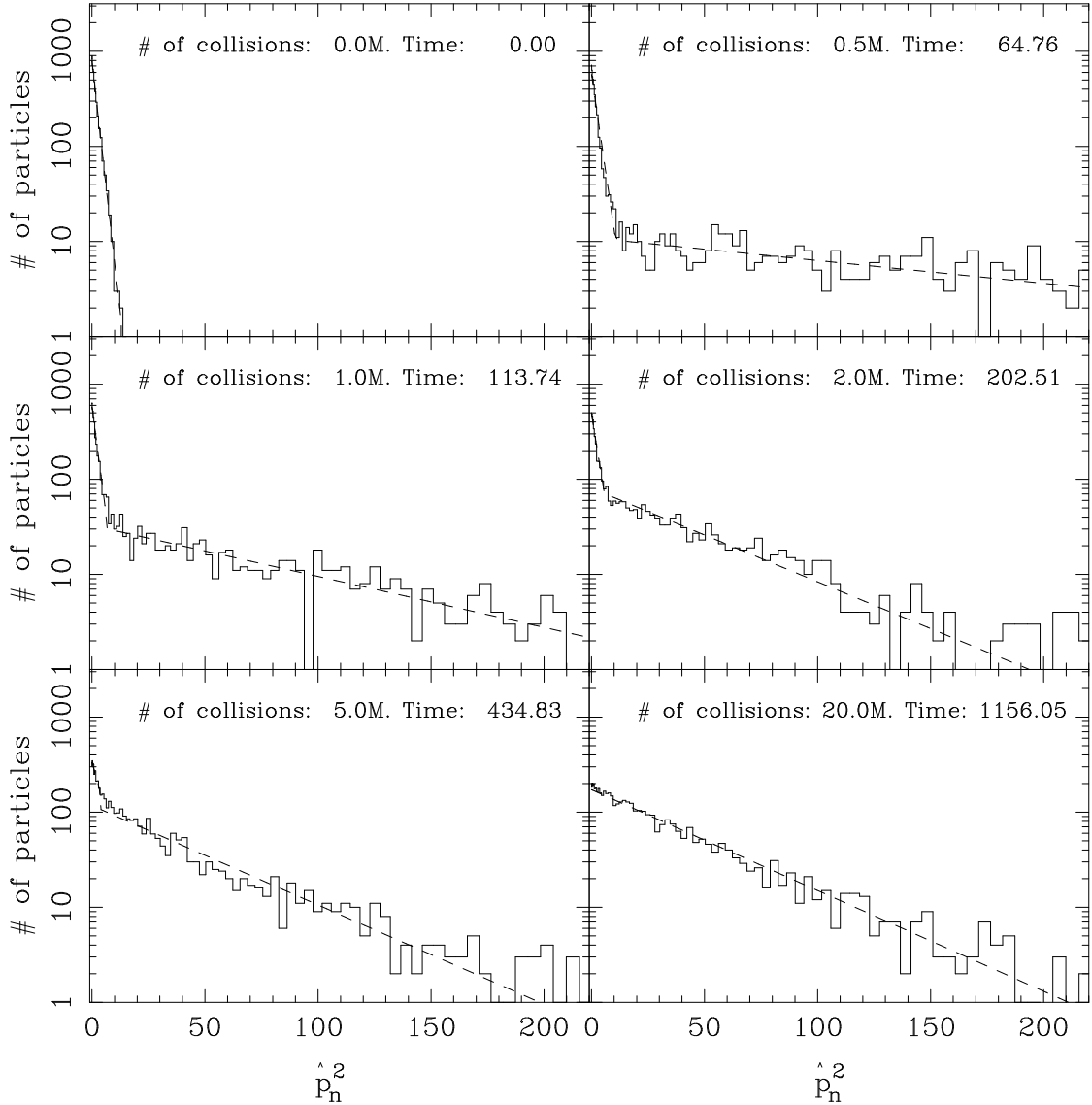}
\caption{\label{fig:mono}Results of the numerical
simulations of an isolated  monodimensional ideal gas
when it relaxes towards equilibrium  (lower right) from
an initial condition in which two very energetic
particles are introduced into the gas in equilibrium (upper left).
Histograms of the one-particle momentum
distribution with a log vertical axis
at various times are shown as solid
lines. A double Gaussian fit, the maximum complexity
distribution, is shown as dashed lines.}
\end{figure}

In more detail, and using arbitrary units from now on,
the gas consisted of 10\,000 pointlike particles
colliding with each other elastically.
The particles were positioned with 
alternating masses of $1$ and $2$ on regular intervals 
on a linear space
10\,000 units long. The system has no boundaries,
i.e., the last particle in this linear
space was allowed to collide with the first one, in a way similar
to a set of rods on a circular ring.
Two distinct masses in the system
were used because 
a monodimensional gas can thermalize only if
its constituent particles have at least two
different masses.
Initially 9998 particles were given initial conditions
following a Gaussian distribution with mean zero
velocity and a mean energy of $1/2$, giving a
total mean energy for the system of nearly $5000$.
These particles where then allowed to undergo $20$ million 
collisions in order for the system to reach the initial state of 
equilibrium, i.e., a Gaussian distribution.
After that, at time $t=0$ two extremely energetic particles
of mass $1$ and $2$ are introduced at two neighboring points, 
such that the total system
has zero momentum and an energy of $150\,000$.
The system then undergoes another $20$ million collisions 
to reach again the equilibrium Gaussian distribution
after a total elapsed time of $\Delta t = 1156.05$.
We record the time evolution of the one-dimensional momentum distribution
in Fig. \ref{fig:mono}, where the square of the
generalized particle momentum is given by the
variable $\hat{p}^2_i \equiv p^2_i / m_i$,
with $p_i$ and $m_i$ the momentum and mass of particle $i$.
The theoretical maximum complexity 
distribution (derived below) for this system is approximated by
two non-overlapping Gaussian distributions and is also fitted 
as dashed lines in Fig. \ref{fig:mono}.
Let us remark that the system stays
in this double Gaussian, the maximum complexity distribution, 
during a large part of its out of equilibrium phase.
The two clearly visible slopes of Fig. \ref{fig:mono} 
are related with the two different mean energies
associated with both Gaussian distributions.
As the system approaches equilibrium, both Gaussian distributions
merge into one.

\begin{figure}
\includegraphics[angle=-90,width=0.9\columnwidth]{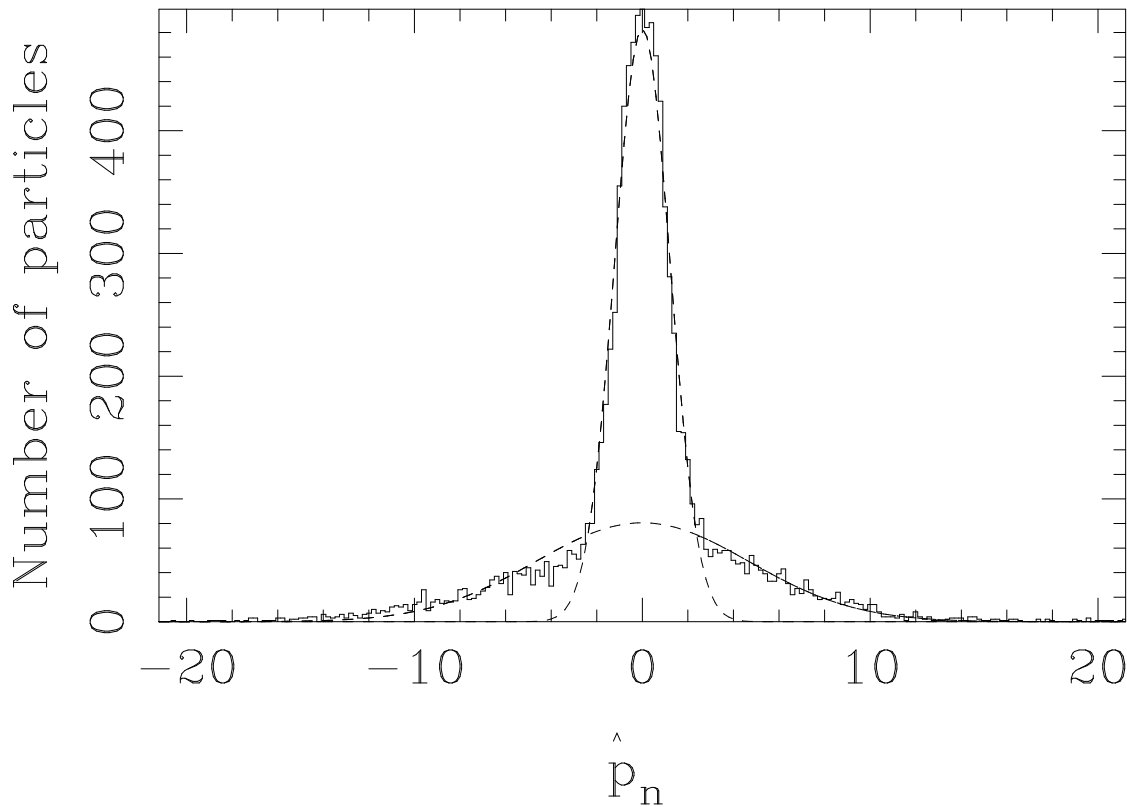}
\caption{\label{fig:monogaus}Numerical
simulations of an isolated  monodimensional ideal gas
as it relaxes to equilibirum at a particular time,
202.5 after 2 million collisions.
The histogram of the one particle momentum distribution
is shown as a solid line. The two fitted gaussian
distributions, the maximum complexity distribution,
are shown as dashed lines.}
\end{figure}

In Fig. \ref{fig:monogaus} the momentum
distribution is shown at a particular moment
out of equilibrium (solid line). In this case a direct histogram 
of the momentum distribution is shown.
The double gaussian or maximum
complexity distribution is clearly seen.
In this Figure, the complete gaussian
distributions are shown for clarity (dashed lines), but the
fit to the numerical simulated data
has been done with non-overlaping Gaussian
functions.

\section{Maximum Complexity distribution}

\begin{figure}
\includegraphics[angle=-90,width=0.9\columnwidth]{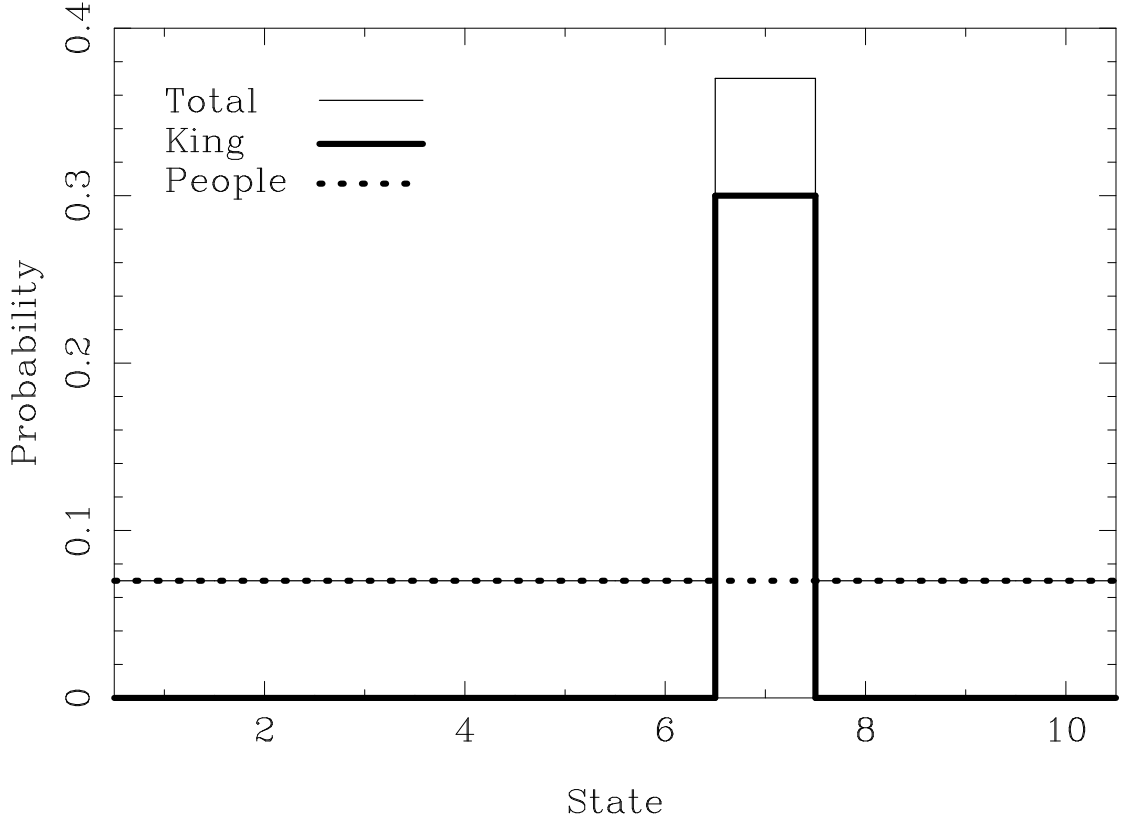}
\caption{\label{fig:maxcomp}
Microcanonical maximum complexity distribution derived
in Ref. \cite{calbet01} (thin solid line). 
It is the sum of the king (thick solid line) 
and the people, or equiprobability, distribution (dotted line).}
\end{figure}

We now sketch the analytical derivation of the
 microcanonical maximum complexity distribution
for a system with a huge number of accessible states.
In Ref. \cite{calbet01}, 
the maximum complexity distribution
for an isolated system with a discrete number of accessible states
was derived. This type of distribution was found to be important 
in the path towards equilibrium for a particular isolated gas, 
the tetrahedral gas. When this system is out equilibrium it decays to
equilibrium by approaching the trajectory formed by all the 
maximum complexity distributions and called the maximum complexity path.
Finding these extremal distributions requires 
solving a variational problem.
The {\it complexity} $C$ (see Ref. \cite{lopez-ruiz95}), is defined as

\begin{equation}
C = D \cdot H,
\end{equation}

where the {\it disequilibium}, $D$, is defined as the distance to the
microcanonical equilibrium probability distribution, 
the equiprobability,
and $H$ is the {\it normalized entropy},

\begin{equation}
\label{eq:h}
D = \sum_{i=1}^N ( f_i - 1/N )^2 \; {\rm and} \;
 H=-(1/\ln N) \sum_{i=1}^N f_i \ln f_i,
\end{equation}

where $N$ is the number of accessible states and $f_i$,
with $i=1,2,\ldots,N$, are the probabilities of permanence 
that the system presents for the different discrete accessible states $i$.
Thus, the microcanonical maximum complexity distribution 
can be derived by finding the maximum disequilibrium
for a given entropy using Lagrange multipliers \cite{calbet01}. 
The results are shown in Table \ref{tab:maximum}. 
Note that, for a given entropy, maximizing the disequilibrium
is equivalent to minimizing
the Tsallis entropy with parameter $q=2$.
Note also that for an isolated
system the entropy variable is equivalent to a streched
time scale, due to its monotonic increase with time
given by the second law of thermodynamics.
The results of Table \ref{tab:maximum}, 
graphically represented in
Fig. \ref{fig:maxcomp}, show that the maximum complexity
distribution can be split into two components.
One of them consists of a background 
equiprobability distribution for all accessible states, 
which will be denoted as the ``people distribution''.
The other one, with the remaining probability,
comprises the particular state with the highest probability 
and is called the ``king distribution''.
The final maximum complexity distribution is
the sum of the people and the king distributions. 
When the system reaches the equilibrium, the king
and people distributions merge leaving only the equiprobability
distribution.

\begin{table}
\caption{\label{tab:maximum} Probability values, $f_j$, 
that give a maximum
of disequilibrium, $D$, or equivalently complexity,
for a given entropy, $H$.}
\begin{ruledtabular}
\begin{tabular}{ccc}
Number of states with $\displaystyle f_j$
    & $\displaystyle f_j$ & Range of $\displaystyle f_j$\\
\hline
$\displaystyle 1$         & $f_{\rm max}$
         & $\displaystyle 1/N\ \ldots\ 1$\\
$\displaystyle N - 1$     & $(1 - f_{\rm max})/(N - 1)$ & 
         $\displaystyle 0\ \ldots\ 1/N$\\
\end{tabular}
\end{ruledtabular}
\end{table}

\section{Maximum Complexity distribution of an isolated monodimensional
ideal gas}

The derivation of a maximum complexity
distribution will be based on the symmetry of the 
momentum phase space and on particular
initial condition considerations.
As an extension of the former discrete case given in Table \ref{tab:maximum},
the distribution we are looking for will have two components.
The people distribution component will be the equiprobability
distribution and the king distribution will be made by choosing 
a particular dominant state according to plausible
arguments derived from the initial state of the system.

First, let us obtain by symmetry arguments how the functional dependence
of the equilibrium distribution, or equivalently the equiprobabilty one,
 looks when only one variable
is maintained and all the rest are integrated. 
In an isolated ideal gas with $n$ particles, 
all accessible states lie on the surface 
of a hypersphere in the $\hat{p}_n$-momentum phase space.
If the energy of a single particle is
$e_i= \hat{p}_i^2 / 2$, then the total
energy of the ensemble is $E=\sum_{i=1}^n e_i$.
The mean energy per particle $e$ will be $e=E/n$.
If the gas is in equilibrium, the microcanonical distribution, 
$h$, is the equiprobability for all accessible states,
i.e., all points in phase space lying on the hypersphere surface
have the same weight in the distribution $h$. This distribution 
is given by the expression
\begin{eqnarray}
h(\theta_{n-1},\theta_{n-2},\dots,\theta_1) 
d \theta_{n-1} d \theta_{n-2} \dots d \theta_1 = \nonumber\\
r^{n-1} d \theta_{n-1} \sin \theta_{n-1} d \theta_{n-2}
\sin \theta_{n-1} \sin \theta_{n-2} d \theta_{n-3} \dots \nonumber\\
\sin \theta_{n-1} \sin \theta_{n-2} \dots \sin \theta_2 d \theta_1,
\end{eqnarray}
where the original phase space variables,
$(\hat{p}_1,\hat{p}_2,\dots,\hat{p}_n)$,
have been converted to the spherical coordinates,
$(r,\theta_1,\theta_2,\dots,\theta_{n-1})$,
with $r^2 / 2 = n e$.
To obtain the one-particle momentum distribution
of this system in equilibrium
$h$ can be integrated 
over all coordinates except $\theta_{n-1}$ obtaining
the function $g$,

\begin{equation}
g(\theta_{n-1}) d \theta_{n-1}=
C' r^{n-1} \sin^{n-2} \theta_{n-1} d \theta_{n-1},
\end{equation}

where $C'$ is the constant of integration.
Converting this result back to the $\hat{p}_n$-momentum coordinate
via the relation $\hat{p}_n = r \cos \theta_{n-1}$
we have the final distribution $f$,

\begin{equation}
f(\hat{p}_n) d \hat{p}_n = C' r^{n-2} \left( 1 - \hat{p}_n^2/r^2
\right)^
\frac{n-3}{2} 
d \hat{p}_n.
\end{equation}

Taking the limit for a large number, $n$, of particles
the one-particle momentum distribution in equilibrium is obtained,

\begin{equation}
f_{EQ}(\hat{p}_n) d \hat{p}_n = C \exp(-\hat{p}_n^2/4 e) d \hat{p}_n,
\end{equation}

where $C=\sqrt{1/(4\pi e)}$ is the normalization constant.
Substituting $e= K T / 2$, $K$ being the Boltzmann constant
and $T$ the temperature, the familiar
one-particle ideal-gas momentum distribution
is obtained, i.e., the Maxwell--Boltzmann distribution. 
It is remarkable that this result has been obtained 
in the microcanonical ensemble although this distribution is
usually presented as a typical derivation from the 
canonical formalism. 
The people distribution, $f_P$, being an equiprobability distribution,
its
formal expression will be the same as the equilibrium one,

\begin{equation}
f_P(\hat{p}_n) =C_P \exp(-\hat{p}_n^2/4 e_P).
\end{equation}

\begin{figure}
\includegraphics[angle=-90,width=0.9\columnwidth]{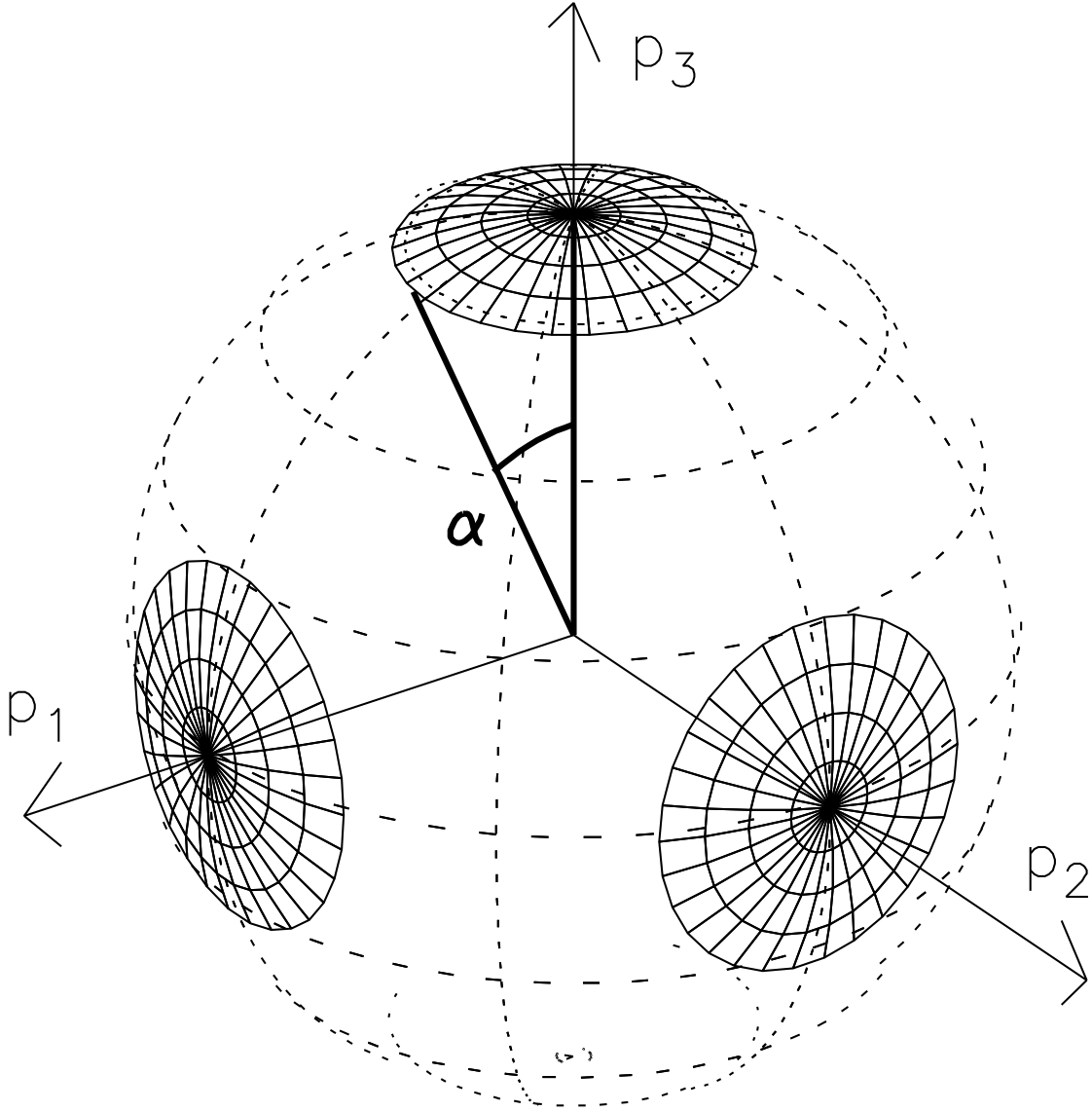}
\caption{\label{fig:king} Non-zero constant probability surfaces of the 
people (dashed mesh) and 
king (solid mesh) distributions on the space ($3D$ sphere) 
of accessible states of a three particle monodimensional ideal gas.}
\end{figure}

Fig. \ref{fig:king} shows the people distribution
in a dimensional space with just three particles
as a dashed mesh surface.

Second, let us discuss how to find a 
king distribution for our system. 
In the scenario of injecting two extremely energetic
particles into the gas in equilibrium, the states with the
highest velocities will be populated with a certain number
of particles in phase space.
 Using symmetry considerations,
all high-velocity components in phase space for each one of those particles
should be equivalent in this distribution. 
It then seems plausible to assume that the high-probability state 
of the king distribution can be formed by all the spherical caps in phase space
centered on the maximum possible velocities for each particle. 
Fig. \ref{fig:king} illustrates this concept
in a small dimensional space with just three particles. 
The king distribution in this phase space is shown as a mesh
of solid lines: the probability of a state lying inside
one of these caps is uniform and zero if it lies outside.
Each of these caps is identified by the angle $\alpha$ shown
in the figure. Integrating  for all momentum dimensions except one, 
the one particle momentum distribution $f_K$ is obtained,

\begin{equation}
f_K(\hat{p}_n) = \quad
\begin{cases}
\Phi(\hat{p}_n)  & \mathrm{if} \; \hat{p}_n <  r\sin \alpha,\\
0  & \mathrm{if} \;  r \sin \alpha \leq \hat{p}_n < r \cos \alpha,\\
 C_K \; {\rm e}^{(-\hat{p}_n^2/4 e_K)} &
 \mathrm{if} \;  r\cos \alpha \leq \hat{p}_n,
\end{cases}
\end{equation}

Since the spherical caps that are over the $\hat{p}_n = 0$ point only 
cover a part of the hypersphere surface given by the set 
of points $(\hat{p}_1,\hat{p}_2,\dots,\hat{p}_{n-1},\hat{p}_n\simeq 0)$, 
the function $\Phi(\hat{p}_n)$ will satisfy,
$\Phi(\hat{p}_n) < C_K \exp(-\hat{p}_n^2/4 e_K)$.

The final exact expression for the maximum complexity
non-equilibrium distribution will be the sum of both distributions,

\begin{equation}
f_{MC}(\hat{p}_n) = f_K(\hat{p}_n) + f_P(\hat{p}_n).
\end{equation}

We can simplify this equation by noting that,
for the $\hat{p}_n <  r\sin \alpha$ cases,
$\Phi(\hat{p}_n) \ll f_P(\hat{p}_n)$,
and that, for the $r \cos \alpha < \hat{p}_n$ cases,
$f_P(\hat{p}_n) \ll f_K(\hat{p}_n)$.

As a first approximation, the maximum 
complexity non-equilibrium distribution is obtained:

\begin{equation}
\label{eq:aproxmaxcomp}
f_{MC}(\hat{p}_n) \cong \quad
\begin{cases}
C_P \exp(-\hat{p}_n^2/4 e_P) & \mathrm{if} \; \hat{p}_n < \hat{p}_0,\\
 C_K \exp(-\hat{p}_n^2/4 e_K)
& \mathrm{if} \; \hat{p}_0 \leq \hat{p}_n,
\end{cases}
\end{equation}

where the momentum $\hat{p}_0 \equiv  r\cos \alpha$ is defined.
Hence the final maximum complexity distribution is a 
two-component non-overlapping Gaussian function
characterized by the parameters $e_P$ and $e_K$, which vary 
with time when the system decays towards equilibrium,
as seen in Fig. \ref{fig:mono}.

\section{Conclusion}

In conclusion, a maximum complexity 
one-particle momentum distribution has been derived
for an isolated monodimensional
ideal gas far from equilibrium. It is based on maximizing the
disequilibrium, or equivalently minimizing the Tsallis
entropy with parameter $q=2$, 
for a given entropy, or equivalently time, in
an isolated system. In a first approximation,
the maximum complexity distribution is 
a double non-overlapping Gaussian distribution.
Numerical simulations of a particular isolated monodimensional
gas show, in clear agreement with our analytical result,
a double Gaussian distribution when it decays towards equilibrium.

\begin{acknowledgments}
The figures in this paper have been prepared using the numerical
programming language PDL (http://pdl.perl.org).
\end{acknowledgments}

\bibliography{manuscript}

\begin{thebibliography}{9}
\expandafter\ifx\csname natexlab\endcsname\relax\def\natexlab#1{#1}\fi
\expandafter\ifx\csname bibnamefont\endcsname\relax
  \def\bibnamefont#1{#1}\fi
\expandafter\ifx\csname bibfnamefont\endcsname\relax
  \def\bibfnamefont#1{#1}\fi
\expandafter\ifx\csname citenamefont\endcsname\relax
  \def\citenamefont#1{#1}\fi
\expandafter\ifx\csname url\endcsname\relax
  \def\url#1{\texttt{#1}}\fi
\expandafter\ifx\csname urlprefix\endcsname\relax\def\urlprefix{URL }\fi
\providecommand{\bibinfo}[2]{#2}
\providecommand{\eprint}[2][]{\url{#2}}

\bibitem[{\citenamefont{Tsallis}(1988)}]{tsallis88}
\bibinfo{author}{\bibfnamefont{C.}~\bibnamefont{Tsallis}}, \bibinfo{journal}{J.
  Stat. Phys.} \textbf{\bibinfo{volume}{52}}, \bibinfo{pages}{479}
  (\bibinfo{year}{1988}).

\bibitem[{\citenamefont{Ferri et~al.}(2005)\citenamefont{Ferri, Martinez, and
  Plastino}}]{ferri05}
\bibinfo{author}{\bibfnamefont{G.~L.} \bibnamefont{Ferri}},
  \bibinfo{author}{\bibfnamefont{S.}~\bibnamefont{Martinez}}, \bibnamefont{and}
  \bibinfo{author}{\bibfnamefont{A.}~\bibnamefont{Plastino}},
  \bibinfo{journal}{J. Stat. Mech.} \textbf{\bibinfo{volume}{4}},
  \bibinfo{pages}{P04009} (\bibinfo{year}{2005}).

\bibitem[{\citenamefont{Wilk and Wlodarczyk}(2000)}]{wilk00}
\bibinfo{author}{\bibfnamefont{G.}~\bibnamefont{Wilk}} \bibnamefont{and}
  \bibinfo{author}{\bibfnamefont{Z.}~\bibnamefont{Wlodarczyk}},
  \bibinfo{journal}{Phys. Rev. Lett.} \textbf{\bibinfo{volume}{84}},
  \bibinfo{pages}{2770} (\bibinfo{year}{2000}).

\bibitem[{\citenamefont{Boghosian}(1996)}]{boghosian96}
\bibinfo{author}{\bibfnamefont{B.}~\bibnamefont{Boghosian}},
  \bibinfo{journal}{Phys. Rev. E} \textbf{\bibinfo{volume}{53}},
  \bibinfo{pages}{4754} (\bibinfo{year}{1996}).

\bibitem[{\citenamefont{Reynolds}(2003)}]{reynolds03}
\bibinfo{author}{\bibfnamefont{A.}~\bibnamefont{Reynolds}},
  \bibinfo{journal}{Phys. Fluids} \textbf{\bibinfo{volume}{15}},
  \bibinfo{pages}{L1} (\bibinfo{year}{2003}).

\bibitem[{\citenamefont{Beck and Cohen}(2003)}]{beck03}
\bibinfo{author}{\bibfnamefont{C.}~\bibnamefont{Beck}} \bibnamefont{and}
  \bibinfo{author}{\bibfnamefont{G.}~\bibnamefont{Cohen}},
  \bibinfo{journal}{Physica A} \textbf{\bibinfo{volume}{322}},
  \bibinfo{pages}{267} (\bibinfo{year}{2003}).

\bibitem[{\citenamefont{Gell-Mann}(1995)}]{gell-mann95}
\bibinfo{author}{\bibfnamefont{M.}~\bibnamefont{Gell-Mann}},
  \bibinfo{journal}{Complexity} \textbf{\bibinfo{volume}{1}},
  \bibinfo{pages}{16} (\bibinfo{year}{1995}).

\bibitem[{\citenamefont{Calbet and Lopez-Ruiz}(2001)}]{calbet01}
\bibinfo{author}{\bibfnamefont{X.}~\bibnamefont{Calbet}} \bibnamefont{and}
  \bibinfo{author}{\bibfnamefont{R.}~\bibnamefont{Lopez-Ruiz}},
  \bibinfo{journal}{Phys. Rev. E} \textbf{\bibinfo{volume}{63}},
  \bibinfo{pages}{066116(9)} (\bibinfo{year}{2001}).

\bibitem[{\citenamefont{Lopez-Ruiz et~al.}(1995)\citenamefont{Lopez-Ruiz,
  Mancini, and Calbet}}]{lopez-ruiz95}
\bibinfo{author}{\bibfnamefont{R.}~\bibnamefont{Lopez-Ruiz}},
  \bibinfo{author}{\bibfnamefont{H.}~\bibnamefont{Mancini}}, \bibnamefont{and}
  \bibinfo{author}{\bibfnamefont{X.}~\bibnamefont{Calbet}},
  \bibinfo{journal}{Physics Letters A} \textbf{\bibinfo{volume}{209}},
  \bibinfo{pages}{321} (\bibinfo{year}{1995}).

\end{thebibliography}

\end{document}